\documentclass[letterpaper,journal,twoside,12pt,draftcls,onecolumn]{IEEEtran}
\pdfoutput=1
\IEEEoverridecommandlockouts
\usepackage{epsfig}
\usepackage{amsmath}
\usepackage{theorem}
\usepackage{amsmath}
\usepackage{mathrsfs}
\usepackage{amsfonts}
\usepackage{amssymb}
\usepackage{mathrsfs} 
\usepackage{euscript}
\usepackage{graphicx}
\usepackage{epstopdf}
\usepackage{multirow}
\usepackage{algorithm}
\usepackage{algorithmic}
\usepackage{cite}
\usepackage{url}
\usepackage{color}

\begin{document}
\title{Joint User Association and Resource Allocation in the Uplink of Heterogeneous Networks\\
{\footnotesize \textsuperscript{}}
\vspace{-10mm}
}
\vspace{-10mm}
\author{
\IEEEauthorblockN{Ata Khalili,~\textit{Student Member,~IEEE},~Soroush Akhlaghi,~Hina Tabassum,~\textit{Senior Member,~IEEE},~and 
Derrick Wing Kwan Ng,~\textit{Senior Member,~IEEE}}
\vspace{-10mm}
\thanks{ A.~Khalili and S.~Akhlaghi are with Department of Engineering, Shahed University,~Tehran,~Iran
(e-mail: ata.khalili@ieee.org,~akhlaghi@shahed.ac.ir).
~H.~Tabassum is with the Department of Electrical Engineering and Computer Science at York University, Canada
(e-mail:hina@eecs.yorku.ca).~D. W. K. Ng is with the School of Electrical Engineering
and Telecommunications, University of New South Wales, Sydney,
NSW 2052, Australia (e-mail: w.k.ng@unsw.edu.au).~H. Tabassum is supported by funding from the Natural Sciences and Engineering Research (NSERC) Discovery Program. ~D. W.
K. Ng is supported by funding from the UNSW Digital Grid Futures Institute,
UNSW, Sydney, under a cross-disciplinary fund scheme and by the Australian
Research Council’s Discovery Project.}}
\maketitle

\begin{abstract}
\textcolor{black}{This letter considers the problem of joint user association (UA),~sub-channel assignment,~antenna selection~(AS),~and power control in the uplink~(UL)~of a heterogeneous network such that the data rate  of small cell users can be maximized while the \textcolor{black}{macro-cell users are protected by imposing a threshold on the cross-tier interference}}. The considered problem is a non-convex mixed integer non-linear programming~(MINLP).~To tackle the problem,~we decompose the original problem into two sub-problems: (i) joint UA,~sub-channel assignment,~and AS, and (ii) power control. Then, we iteratively solve the sub-problems  by applying the tools from majorization-minimization~(MM) theory and augmented Lagrange method, respectively, and obtain locally optimal solutions for each sub-problem.\textcolor{black}{~Simulation results illustrate that our proposed scheme outperforms existing schemes.}
Complexity analysis of the proposed algorithm is also presented.
\end{abstract}
%
\IEEEpeerreviewmaketitle
\vspace{-8mm}
\section{Introduction}
\vspace{-2mm}
{Radio resource allocation design is indispensable in mitigating network interference,~improving achievable data rates, and in turn avoiding the under-utilization of resources~\cite{Dr.Tabassum,3}}.
For instance, the problem of joint sub-channel assignment
and power control was considered in \cite{3} to maximize
the network throughput. The original problem was converted
into a standard form of difference of convex functions (DC)
programming and a sub-optimal solution was developed based
on successive convex approximation.

{Furthermore, a plethora of research studies considered user association~(UA), sub-channel and power allocation problems in  downlink heterogeneous networks~(HetNets)~\cite{massive,multi,Poor, Contract}}.
In \cite{multi}, joint resource allocation and UA  was solved for downlink HetNets to maximize the weighted sum-rate.~The authors decomposed the original problem into a series of sub-problems. In the first sub-problem, the joint UA and sub-channel allocation for fixed power allocation was solved based on the bipartite matching problem and then for the chosen assignment the power control was solved through DC method.~The problem of joint UA and resource allocation was  investigated in \cite{massive} for the downlink HetNets  to maximize the alpha fairness network utility. The solution was based on the Lagrangian dual analysis resulting in a sub-optimal solution.~Besides,~the problem of UA and power allocation in mmWave was investigated in \cite{Poor}. This problem was iteratively solved by relaxing the integer variable and Lagrangian dual decomposition.~\textcolor{blue}{The authors in \cite{Contract} proposed a contract-based traffic offloading and resource allocation mechanism for the software-defined  ultra-dense heterogeneous wireless networks.}

Along another note, multiple-input multiple-output (MIMO) increases the reliability and capacity of wireless networks. Nevertheless, due to the high cost and complexity involved in the pre-processing and post-processing at the transmitter and receivers, appropriate antenna selection (AS) schemes are needed to reduce the hardware complexity and overheads~\cite{LTE,Beyound}. AS schemes are incorporated in the uplink of long-term evolution advanced~(LTE-A) to minimize implementation complexity and feedback information compared to other beamforming/precoding techniques \cite{LTE}.~A relevant study is \cite{Dr. Kwan} where the resource allocation and AS was considered to minimize the total power consumption.

{To the best of our knowledge,~the problem of joint UA,~AS,~sub-channel,~and power allocation in the uplink of a two-tier HetNet has not been investigated in the literature. {In uplink works including [2],~[7],~[11], transceivers are equipped with single antenna.~Furthermore,~in \cite{3}, only sub-channel assignment and power control was considered,~Moreover,~the study in \cite{globecom2018} was focused on devising a heuristic joint uplink UA and resource allocation scheme to minimize users' transmit power subject to quality-of-service (QoS) constraints.}}

{In this letter,~we aim to bridge gap.~In particular,~we consider the UA,~AS,~sub-channel and power allocation in the uplink of a two-tier HetNet to maximize the network data rate. The considered problem is a non-convex  mixed integer nonlinear programming (MINLP), therefore, we decouple the original problem into two sub-problems and solve them iteratively until \textcolor{black}{the convergence to a suboptimal solution is achieved.}~Thus,~the main contributions of this paper are summarized as follows:\\
$\bullet$~We propose a novel method to cope up with the multiplication of two binary variables based on majorization minimization approach by constructing a sequence of surrogate function and also applying the abstract Lagrangian duality which yields an efficient locally optimal solution.\\
$\bullet$~We propose an efficient power control policy based on the Augmented Lagrange method (ALM) to obtain a locally optimal and less complex solution. Note that ALM outperforms the traditional sub-gradient or dual-descent methods.\\  
$\bullet$ Finally, we compare the performance of the proposed resource allocation algorithm with existing schemes and present a complexity analysis of the algorithm.}
\vspace{-5mm}
\section{System Model and Problem Formulation}\label{sys_mod}
\vspace{-1mm}
We consider a heterogeneous orthogonal frequency division multiple access (OFDMA)-based network where a macro-cell (MC) shares its sub-channels with $B$ open access small cells (SCs)\footnote{All BSs are equipped with multiple antennas.~However, AS at BSs is out of scope of this work and is assumed to be predefined at the BSs, i.e., each antenna is reserved for a subset of users.}. The set of BSs is represented by~$\mathcal{B}=\{{0,1,2,...,B}\}$, where~$0\in\mathcal{B}$ denotes the MC's index.~${I}$ denotes the total number of users and the set of SC users which is connected to the BS $b$ is denoted by $\mathcal{I}_{b}$. Users are randomly distributed among cells and each user is equipped with $\mathcal{A}=\{1,2,...,A\}$ antennas.~$\mathcal{M}=\{{1,2,...,M}\}$ is the fixed set of available sub-channels at each~cell.~$h_{i,{b}}^{m a}$ denotes the~UL~channel coefficients from the $i^{\mathrm{th}}$ user over the $m^{\mathrm{th}}$ sub-channel when the $a^{\mathrm{th}}$ antenna is selected.~\textcolor{black}{We assume that perfect channel state information (CSI) is available at a centralized resource allocator to design resource alloction schemes}\footnote{\textcolor{black}{We assume that each BS broadcasts some pilot signals to users. Next, each user estimates the CSI and sends it to the related BS via a feedback channel.~Then, all BSs send the CSI to the centralized controller for resource allocation.~In particular,~each BS  sends some orthogonal preambles in the downlink  to the users and obtains the CSI by listening to the sounding reference signals transmitted by the users.}}.~Let~$p_{ib}^{m a}$ denote the transmit power of the $i^{{\mathrm{th}}}$~user to the BS $b$~over the $m^{{\mathrm{th}}}$~sub-channel when the $a^{{\mathrm{th}}}$ antenna of this user being selected.~Furthermore,~$s_{ib}^{m}$ indicates BS $b$ allocates sub-channel $m$ to user $i$,~and~$x_{i}^{m a}$~represents the AS variable of the $i^{{\mathrm{th}}}$ user over the $m^{{\mathrm{th}}}$ sub-channel from the $a^{\mathrm{th}}$ antenna of this user. When the $a^{{\mathrm{th}}}$ antenna is selected of user $i$ who is associated to the $b^{{\mathrm{th}}}$ BS, the data rate over the $m^{{\mathrm{th}}}$ sub-channel  is:
\textcolor{black}{\begin{align}\label{rr}
  R^{m a}_{ib}= \log_2\!\!\bigg(\!1\!+\!\frac{ p^{m a}_{ib} |h^{m a}_{i{,b}}|^2}{\sigma^2+\underset{b'\in\mathcal{B}\backslash\{b\}}{\overset{}{\mathop \sum }}~\underset{l\neq i}{\overset{}{\mathop \sum }}~\underset{a'\in \mathcal{A}}{\overset{}{\mathop \sum }} s^{m}_{ib'}\ x^{ma'}_{i}p^{m{a'}}_{lb'} |h^{m a'}_{l{,b}}|^2}\!\bigg),\end{align}}\textcolor{black}{where~$\sigma^2$ is the additive noise power and $\mathcal{A}\setminus B $ denotes the set whose elements are in $\mathcal{A}$ and not in $\mathcal{B}$.~Furthermore,} $\underset{b'\in\mathcal{B}\backslash\{b\}}{\overset{}{\mathop \sum }}~\underset{l\in\mathcal{I}_{b'}}{\overset{}{\mathop \sum }} \underset{a'\in \mathcal{A}}{\overset{}{\mathop \sum }} s^{m}_{ib'}\ x^{ma'}_{i}p^{m{a'}}_{lb'} |h^{m a'}_{l{,b}}|^2$ is the co-channel interference. Our objective is to maximize the total UL throughput of SC users while optimizing sub-channel,~UA,~AS, and power allocation. The optimization problem can be written as follows
  \vspace{-2mm}
\begin{align}\label{max_prob.1}
\max_{\textbf{x},\textbf{s},\textbf{p}}&\underset{b\in\mathcal{B}\backslash\{0\}}{\overset{}{\mathop \sum }}~\underset{i\in \mathcal{I}_{b} }{\overset{}{\mathop \sum }}~\underset{m\in \mathcal{M}}{\overset{}{\mathop
\sum }}\underset{a\in\mathcal{A}}{\overset{}{\mathop \sum }} x^{m a}_{i}s^{m}_{ib}\ R^{m a}_{ib}\nonumber\\
\textrm{s.t.}~ &{{C}_{1}}:\sum_{b\in \mathcal{B}\backslash\{0\}}\underset{a\in\mathcal{A}}{\overset{}{\mathop \sum }}\,\underset{m\in \mathcal{M}}{\overset{}{\mathop \sum }}\,x_{i}^{m a}s_{ib}^{m}p_{ib}^{m a}\le {{p}_{\max}},\nonumber\\
 &{{C}_{2}}:\underset{b\in\mathcal{B}\backslash\{0\}}{\overset{}{\mathop \sum }}\,\underset{i\in \mathcal{I}_{b}}{\overset{}{\mathop \sum }}\underset{a\in\mathcal{A}}{\overset{}{\mathop \sum }}\ x_{i}^{m a}s_{ib}^{m}p_{ib}^{m a}h^{m a}_{i_{b,0}}\le {I}^{m}_{\textrm{th}},\nonumber\\
 &\textcolor{black}{{{C}_{3}}:\underset{a\in\mathcal{A}}{\overset{}{\mathop \sum }}\,\underset{m\in \mathcal{M}}{\overset{}{\mathop \sum }}\,x_{i}^{m a}s_{ib}^{m}R^{m a}_{i_{b}}\geq {R}_{\textrm{min}}},\nonumber\\
~&{{C}_{4}}:p_{ib}^{m a}~\ge 0,~{{C}_{5}}:\underset{i\in \mathcal{I}_{b} }{\overset{}{\mathop \sum }}\,s_{ib}^{m}\leq1,\nonumber\\
~&{{C}_{6}}:\underset{m\in \mathcal{M}}{\overset{}{\mathop
\sum }}~\underset{b\in \mathcal {B} }{\overset{}{\mathop \sum }}\,s_{ib}^{m}\leq1,~{{C}_{7}}:\underset{a\in \mathcal{A}}{\overset{}{\mathop \sum }},x_{i}^{m a}=1,~~\nonumber\\
&{{C}_{8}}:~x_{i}^{m a}\in \left\{ 0,1 \right\},~C_9:s_{ib}^{m}\in \left\{ 0,1 \right\}.
\end{align}
The vector of power allocation,~joint sub-channel-BS assignment,~and antenna variables are defined as $\textbf{p}\in \mathbb{R}^{{B}{I} {M} {A}\times1}$,~$\textbf{s}\in \mathbb{Z}^{{B}{I} {M}\times1}$,~and~$\textbf{x}\in \mathbb{Z}^{{M}{I}{A}\times1}$,~respectively.~Constraint~$C_{1}$ in~(\ref{max_prob.1}) indicates that the total transmit power of each user is limited to  $p_{\max}$.~$C_{2}$ imposes a maximum limit of the cross-tier interference arising from small-cell users at the MC,~where ${I}_{\text{th}}$ denotes the maximum tolerable interference level on a given sub-channel $m$ to protect macro-cell users.$C_3$ is the QoS requirement for each user, i.e., a minimum data rate requirement $R_{\mathrm{min}}$ is imposed for each small cell user.~$C_4$~ensures that the allocated transmit power to each user is non-negative.~Constraint~$C_5$~ensures that each sub-channel is assigned to at most one user.~Moreover,~$C_6$ represents that each user is assigned only to one BS.~$C_7$ denotes that each user in each sub-channel makes use of one antenna\footnote{\textcolor{blue}{Our problem formulation and solution are applicable to the case of per-carrier and per-subcarrier depending on the resolution of fast Fourier transform (FFT) /IFFT at the expense of different hardware costs.}}. Finally,~$C_8$ and~$C_9$ indicate that the joint sub-channel-BS indices and the antenna indicators are binary variables.

Joint optimization of the sub-channel,~UA,~AS,~and power allocation is challenging due to non-convexity and the coupling in the objective and constraint.~In general,~such problem is generally intractable.~In order to design an efficient resource allocation, we decompose the original problem into two sub-problems and propose to solve  \eqref{max_prob.1} by  solving the two sub-problems in an iterative manner.~This algorithm is referred as joint power control and scheduling {J-PCS} algorithm.~At the initial point,~the joint UA,~sub-channel assignment,~and AS $\mathbf{s}[0],\mathbf{x}[0]$~are obtained for the initial power allocation $\mathbf{p}[0]$.~Then for the chosen scheduling,~we find the power allocation.~We repeat the process in all subsequent iterations until no further improvement is made.~The corresponding update rule is summarized as follows:
\vspace{-1.5mm}
\begin{align}\label{update}
  &\mathbf{s}[0],\mathbf{x}[0]\rightarrow\mathbf{p}[0]\rightarrow ...\rightarrow  \mathbf{s}[t-1],\mathbf{x}[t-1]\rightarrow\mathbf{p}[t-1]\\ &\rightarrow \mathbf{s}[t],\mathbf{x}[t]\rightarrow\mathbf{p}[t]\rightarrow ...\rightarrow \mathbf{s^*}[t+1],\mathbf{x^*}[t+1]\rightarrow\mathbf{p^*}[t+1].\nonumber
\end{align}
J-PCS algorithm reduces the number of variables by half in each iteration and changes the original problem~\eqref{max_prob.1} into a mathematically tractable form.~Then,~the proposed  algorithm continues until no improvement is achieved in the data rate.
\vspace{-3mm}
\section{Joint Resource Allocation}\label{subch}
\vspace{-1mm}
For a given power~$\textbf{p}^{t-1}$ from the $t-1$ iteration,~the optimization problem in \eqref{max_prob.1} can be simplified to
\vspace{-2mm}
\begin{equation}\label{max_subchannel}
\begin{aligned}
\max_{\textbf{s},\textbf{x}}&\underset{b\in\mathcal{B}\backslash\{0\}}{\overset{}{\mathop \sum }}~\underset{i\in \mathcal{I}_{b} }{\overset{}{\mathop \sum }}~\underset{m\in \mathcal{M}}{\overset{}{\mathop
\sum }}\underset{a\in\mathcal{A}}{\overset{}{\mathop \sum }} x^{m a}_{i}s^{m}_{ib}\ R^{m a}_{ib}(\mathbf{p}^{t-1})\\
\textrm{s.t.}~& {{C}_{1}}-{{C}_{9}}.
\end{aligned}
\end{equation}
The optimization problem \eqref{max_subchannel} is still non-convex due to the product of two binary variables.~For tractability,~we first rewrite the two binary constraints into their equivalent forms \cite{10}:
\vspace{-4mm}
\begin{align}\label{max_prob.4}
&\mathcal{R}_{1}:0\leq x^{m a}_{i}\leq 1,~
\mathcal{R}_{2}:0\leq s^{m}_{ib}\leq 1,\nonumber\\
&\mathcal{R}_{3}:\sum_{i}\sum_{m}\sum_{a} \big(x^{m a}_{i}-(x^{m a}_{i})^2\big) \leq 0,\nonumber\\
&\mathcal{R}_{4}:\sum_{b}\sum_{i}\sum_{m} \big(s^{m}_{ib}-(s^{m}_{ib})^2\big) \leq 0.
\end{align}
Considering~$Q$ as the feasible set spanned by the constraints ${C_1}-{C_7}$, the optimization problem~\eqref{max_subchannel} can be equivalently represented as:
\begin{equation}\label{max_prob.5}
\max_{\textbf{x},\textbf{s}}\ R(\textbf{x},\textbf{s},\textbf{p}^{t-1})\quad \textrm{s.t.}\quad\textbf{x},\textbf{s},\textbf{p} \in Q,\mathcal{R}_1-\mathcal{R}_4,
\vspace{-1mm}
\end{equation}
where,~$R(\textbf{x},\textbf{s},\textbf{p}^{t-1})$ represents the objective function of~\eqref{max_subchannel}. The problem in (\ref{max_prob.5}) is a continuous optimization problem .{~To further facilitate the algorithm design, we adopt the \textit{abstract Lagrangian duality}:
\begin{equation}\label{max_prob.6}
  \max_{\textbf{x},\textbf{s}}\ L(\textbf{x},\textbf{s},\textbf{p}^{t-1},\mu_{1},\mu_{2})\quad \textrm{s.t.}\quad\textbf{x},\textbf{s},\textbf{p} \in Q,~\mathcal{R}_1,\mathcal{R}_2,
\end{equation}
 where~$L(\textbf{x},\textbf{s},\textbf{p},\mu_{1},\mu_{2})$ is the \textit{abstract Lagrangian duality}
associated to~(\ref{max_prob.5}), and is defined as follows} 
\begin{align}\label{max_prob.7}
L(\textbf{x},\textbf{s},\textbf{p},\mu)\triangleq\ &R(\textbf{x},\textbf{s},\textbf{p})-\mu_{1}\sum_{b}\sum_{i}\sum_{m} \Big(s^{m}_{ib}\!-(s^{m}_{ib})^2\Big)\nonumber\\
-&\mu_{2}\sum_{i}\sum_{m}\sum_{a} \Big(x^{m a}_{i}\!-(x^{m a}_{i})^2\Big).
\end{align}
The coefficients~$\mu_i$ for $i=1,2$ are constant values which act as penalty factors. For large values of $\mu_i$,~the optimization~(7) is equivalent to~(6) which means $d^{*}=\min_{\mu_{1},\mu_{2}} \max_{\textbf{x},\textbf{s}}L(\textbf{x},\textbf{s},\textbf{p}^{t-1},\mu_{1},\mu_{2})=\max_{\textbf{x},\textbf{s}} \min_{\mu_{1},\mu_{2}}L(\textbf{x},\textbf{s},\textbf{p}^{t-1},\mu_{1},\mu_{2})=p^{*}$~\cite{10}.
It should be noted that the objective function in~(6)~is still non-convex.~It is straight forward to show that~$x^{a m}_{i}s^{m}_{ib}$~is the same as $\frac{1}{2}\Big(x^{a m}_{i}+s^{m}_{ib}\Big)^{2}-\frac{1}{2}\Big(\big(x^{a m}_{i}\big)^{2}+\big(s^{m}_{ib}\big)^{2}\Big)$.~Now,~we can express (4) as below
\begin{eqnarray}\label{sysmod22}
\begin{aligned}
\max_{\textbf{x},\textbf{s}}~&F(\textbf{x},\textbf{s})-{G(\textbf{x},\textbf{s})}\\
\text{s.t.}~&{{C}_{1}}-C_{9},
\end{aligned}
\end{eqnarray}
\vspace{-3mm}
 where
\begin{equation}\label{functionx}
 F(\textbf{x},\textbf{s})\triangleq\sum_{i,m,a,b}\frac{R^{ma}_{ib}}{2}\Big(x^{a m}_{i}+s^{m}_{ib}\Big)^{2}-\mu_{1}(x^{m a}_{i})-\mu_{2}(s^{m}_{ib}),
 \end{equation}
    \begin{equation}\label{functiong}G(\textbf{x},\textbf{s}) \triangleq \!\!\!\!\!\sum_{i,m,a,b}\frac{R^{ma}_{ib}}{2}\Big(\big(x^{a m}_{i}\big)^{2}+\big(s^{m}_{ib}\big)^{2}\Big) -\mu_{1}(x^{m a}_{i })^{2}-\mu_{2}(s^{m}_{ib})^2.
 \end{equation}
 \normalsize
 Note that in \eqref{functionx} and \eqref{functiong}, we use a short-hand notation $\sum_{i,m,a,b}=\underset{b\in\mathcal{B}\backslash\{0\}}{\overset{}{\mathop \sum }}~\underset{i\in \mathcal{I}_{b} }{\overset{}{\mathop \sum }}~\underset{m\in \mathcal{M}}{\overset{}{\mathop
\sum }}\underset{a\in\mathcal{A}}{\overset{}{\mathop \sum }} $.
Although  both terms in the objective function in (9) are convex,~the optimization problem of (9) is still non-convex~\cite{10,11}. To tackle
the problem, a majorization-minimization approach is
applied by constructing a surrogate function~\cite{11}~using first order Taylor approximation such that  a locally optimal solution can be obtained as
\vspace{-1.5mm}
\begin{eqnarray}\label{main_problem}
\begin{aligned}
\max_{\textbf{x},~\textbf{s}}~~&F(\textbf{x},\textbf{s})-{G(\textbf{x}^{t_{j}-1},\textbf{s}^{t_{j}-1})}\\
&-\nabla_{\textbf{x}}G^{T}(\textbf{x}^{t_{j}-1},\textbf{s}^{t_{j}-1}).(\textbf{x}-\textbf{x}^{t_{j}-1})\\
&-\nabla_{\textbf{s}}G^{T}(\textbf{x}^{t_{j}-1},\textbf{s}^{t_{j}-1}).(\textbf{s}-\textbf{s}^{t_{j}-1})\\
&\text{s.t.}~~{{C}_{1}}-C_{9}.\\
\end{aligned}
\end{eqnarray}
In (12),~$t_{j}$ denotes the iteration number,~$\nabla_{\textbf{s}}$ and $\nabla_{\textbf{x}}$  denote the gradient with respect to~\textbf{s}~and $\textbf{x}$, respectively.~Since the optimization problem in~(\ref{main_problem}) is convex at each iteration, it can be effectively solved using optimization packages incorporating interior-point methods like CVX.
\vspace{-5mm}
\section{Power Control Policy}\label{PA}
\vspace{-2mm}
In this section, we optimize the power allocation given the assigned sub-channels,~BSs,~and antennas. The index $[t-1]$ shows the variables whose values are taken from the previous iteration.~The power allocation problem is stated as:
\vspace{-4mm}
\begin{align}\label{max_prob.2}
\max_{\textbf{p}}&\underset{b\in\mathcal{B}\backslash\{0\}}{\overset{}{\mathop \sum }}~\underset{i\in \mathcal{I}_{b} }{\overset{}{\mathop \sum }}~\underset{m\in \mathcal{M}}{\overset{}{\mathop
\sum }}\underset{a\in\mathcal{A}}{\overset{}{\mathop \sum }} x^{m a}_{ib}[t-1]s^{m}_{ib}[t-1]\ R^{m a}_{ib}\\
\textrm{s.t.}~&{{C}_{1}}:\sum_{b\in \mathcal{B}\backslash\{0\}}\underset{m\in\mathcal{M}}{\overset{}{\mathop \sum }}\,\underset{a\in \mathcal{A}}{\overset{}{\mathop \sum }}\,x_{i}^{m a}[t-1]s_{ib}^{m}[t-1]p_{ib}^{m a}\le {{p}_{\max}},\nonumber\\
&{{C}_{2}}:\underset{b\in\mathcal{B}\backslash\{0\}}{\overset{}{\mathop \sum }}\,\underset{i\in \mathcal{I}_{b}}{\overset{}{\mathop \sum }}\underset{a\in\mathcal{A}}{\overset{}{\mathop \sum }}\ x_{i}^{m a}[t-1]s_{ib}^{m}[t-1]p_{ib}^{m a}h^{m a}_{i_{b,0}}\le {I}^{m}_{\textrm{th}},\nonumber\\
 &{{C}_{3}}:\sum_{b\in \mathcal{B}\backslash\{0\}}\underset{m\in\mathcal{M}}{\overset{}{\mathop \sum }}\,\underset{a\in \mathcal{A}}{\overset{}{\mathop \sum }}\,x_{ib}^{m a}[t-1]s_{ib}^{m}[t-1]R^{m a}_{i_{b}}\geq {R}_{\textrm{min}}.\nonumber
\end{align}
Note that the objective function and~constraint $C_{3}$ are non-convex due to the incorporated interference in the rate function.~As such, there exists a duality gap between the primal and dual problem and to decrease this duality gap, we propose to apply the ALM in which we add  a penalty term to the Lagrange function.~ALM is based on the quadratic penalty function method and it has been shown to outperform traditional  sub-gradient or dual-descent methods.~\textcolor{black}{In contrast to the penalty method, the ALM largely preserves smoothness and no longer requires the existence of a sufficiently large penalty term to guarantee the convergence of the method~\cite{12}.}

Writing the ALM of \eqref{max_prob.2}, we get \eqref{max_prob.6} at the top of the next page (details of the ALM formulation are provided in {\bf Appendix A}),
\begin{figure*}
\begin{align}\label{max_prob.6}
\tiny
&L_{\psi}(\textbf{p},\boldsymbol{\lambda,\mu,\theta})=R(\textbf{p})+\frac{1}{2\psi}\Bigg[\Bigg(\Bigg[\underset{i}{\overset{}{\mathop \sum }}\lambda_{i}+\psi \Big(\underset{b\in\mathcal{B}\backslash\{0\}}{\overset{}{\mathop \sum }}\,\underset{m}{\overset{}{\mathop \sum }}\underset{a}{\overset{}{\mathop \sum }}\,p_{ib}^{ma}- {{p}_{\max}}\Big)\Bigg]^{+}\Bigg)^{2}-\underset{i}{\overset{}{\mathop \sum }}\lambda_{i}^{2} -\underset{i}{\overset{}{\mathop \sum }}\phi_{i}^{2}+\nonumber\\&\Bigg(\Bigg[\sum_{m}\theta_{m}+\psi\bigg( \underset{b\in\mathcal{B}\backslash\{0\}}{\overset{}{\mathop \sum }}\,\underset{i}{\overset{}{\mathop \sum }}\underset{a}{\overset{}{\mathop \sum }}\,p_{ib}^{ma}h_{i,0}^{m a} -{{I}_{\textrm{th}}}\bigg)\Bigg]^{+}\Bigg)^{2}
-
\underset{m}{\overset{}{\mathop \sum }}\theta_{m}^{2}+\Bigg[\underset{i}{\overset{}{\mathop \sum }}\phi_{i}+\psi \Big(\underset{b\in\mathcal{B}\backslash\{0\}}{\overset{}{\mathop \sum }}\,\underset{m}{\overset{}{\mathop \sum }}\underset{a}{\overset{}{\mathop \sum }}\,{{R}_{\min}}-R_{ib}^{ma}\Big)\Bigg]^{+}\Bigg)^{2}\Bigg],\tag{14}
\end{align}
\hrule
\end{figure*}
\normalsize
where $\psi$ is a positive coefficient adjustable penalty parameter and~$(\boldsymbol{\lambda,\theta,\phi})$~are the Lagrange multiplier vectors.~The solution of \eqref{max_prob.6} gives the solution to \eqref{max_prob.2}. Solving \eqref{max_prob.6} is  a two-step procedure. The first step is to maximize the {\em augmented Lagrangian} for the appropriate set and for the second step {\em augmented Lagrangian} as well as penalty parameter are updated.~Based on the above explanation, the sub-gradient method is employed to update the Lagrange multipliers as follows:
\vspace{-3mm}
\begin{align}
\vspace{-4mm}
\nonumber&\lambda^{n+1}_{i}=\Big[\lambda_{i}^{n}+\psi \Big(\underset{b\in\mathcal{B}\backslash\{0\}}{\overset{}{\mathop \sum }}\,\underset{m\in\mathcal{M}}{\overset{}{\mathop \sum }}\underset{a\in\mathcal{A}}{\overset{}{\mathop \sum }}\,p_{ib}^{ma}- {{p}_{\max}}\Big)\Big]^{+},\tag{15}\\\nonumber
&\theta^{n+1}_{m}=\Big[\theta_{m}^{n}+\psi \Big( \underset{b\in\mathcal{B}\backslash\{0\}}{\overset{}{\mathop \sum }}\,\underset{i\in\mathcal{I}_b}{\overset{}{\mathop \sum }}\underset{a\in\mathcal{A}}{\overset{}{\mathop \sum }}\,p_{ib}^{ma}h_{i,0}^{m a} -{{I}_{\textrm{th}}}\Big)\Big]^{+},\tag{16}\\\nonumber
&\phi^{n+1}_{i}=\Big[\phi_{i}^{n}+\psi \Big(\underset{b\in\mathcal{B}\backslash\{0\}}{\overset{}{\mathop \sum }}\,\underset{m\in\mathcal{M}}{\overset{}{\mathop \sum }}\underset{a\in\mathcal{A}}{\overset{}{\mathop \sum }}\,R_{\min}-R_{ib}^{ma}\Big)\Big]^{+},\tag{17}
\end{align}
where the superscript $n$ depicts the iteration number.~Also, the penalty parameter is updated as $\psi^{n+1}=2\psi^{n}$.~The pseudo-code of our proposed solution is presented in \textbf{Algorithm 1.}
\begin{algorithm}[t]
 \small \caption{\small Joint Power Control and Scheduling~(J-PCS)~algorithm}
 \begin{algorithmic}[1]
 \STATE \textbf{Initialize:} $t = 0$,~Set error tolerance $\epsilon =0.1$,~and  $\textbf{p}[0]$.
 \STATE \textbf{while} $|(R^{ma}_{ib})^{(t+1)}-(R^{ma}_{ib})^{(t)}| > \epsilon$  \textbf{do}\\
  \textbf{Step 1.~UA,~Sub-channel assignment,~and AS}
 \STATE \textbf{Initialization for step 1}:~Initialize $t_{j}=0$ and maximum number of iteration $T_{j\max}$, penalty factor $\mu_{1}$,$\mu_{2}\gg1$
  \STATE \textbf{Repeat}
 \STATE \quad  Solve the optimization problem (12) to obtain \textbf{s},~\textbf{x}.
 \STATE Set $t_{j}=t_{j}+1$ and $\textbf{s}^{t_{j}}=\textbf{s}$,~$\textbf{x}^{t_{j}}=\textbf{x}$
 \STATE \textbf{Until} convergence or $t_{j}=T_{j\max}$\\
 \textbf{Step 2.~Power allocation}\\
 \STATE \textbf{while$|(p^{ma}_{ib})^{(n+1)}-(p^{ma}_{ib})^{(n)}| > 10^{-3}$} \textbf{do}
 \STATE Solve the optimization problem (14) with given $\textbf{s}^{*}(t)$,~$\textbf{x}^{*}(t)$ to obtain $\textbf{p}$
 \STATE \quad  update $\lambda^{n+1}_{i}$,~$\theta^{n+1}_{m}$,~and $\phi^{n+1}_{i}$ using (15),~(16),~and (17),~respectively.~Moreover,~update $\psi^{n+1}=2\psi^{n}$
 \STATE   \quad Set $n : = n + 1$.
\STATE \textbf{end while}
\STATE $\textbf{p}^{*}[t+1]=\textbf{p}^{n}$
\STATE Set~$t=t+1$
\STATE \textbf{end while}
\end{algorithmic}
\end{algorithm}
\vspace{-8mm}
\section{Complexity Analysis of the J-PCS Algorithm}
\vspace{-2mm}
Our proposed algorithm is composed of two main sub-problems, i.e., deriving the sub-channel assignment,~UA,~and AS from (12) using CVX~and solving the power allocation based on the ALM.~For the sub-problem, when CVX is adopted,~it employs DC with the interior point method and the number of required iterations is ${\log(C/{t_{j0}\varrho})}/{\log\varepsilon}$~where $C=({B}+{M}+{I}+{B}{I}{M}{A}+{B}{M}+{I}{M}+{I}{B}{M}+{I}{M}{A})$~is the total number of constraints,~$t_{j0}$ is the initial point for approximating the accuracy of interior point method,~$0<\varrho \ll 1$ is the stopping criterion,~and $\varepsilon$ is used to represent the accuracy of the method~\cite{Boyd}.~On the other hand, for the power allocation based on the augmented Lagrangian the order of complexity at each iteration is $\mathcal{O}(IBMA)^{2}$ which is polynomial time~\cite{12}.{\textcolor{black}{~Note the computational complexity of proposed method in [3] is $\mathcal{O}(IBM)^{3}$.}}
\vspace{-5mm}
\section{Simulation Results}\label{sim}
\vspace{-2mm}
In this section, we investigate the performance of the proposed algorithm and compare it to the algorithm proposed in \cite{3}.~\textcolor{black}{The large scale fading of the communication channel is computed according to the path loss formula based on the 3GPP propagation model \cite{Path}, i.e.,  $\text{Path-loss}=PL_{0}+10\theta \log (d)$~(dB),~where $d$ denotes the distance between user and BS,~$PL_{0}$ is the constant path-loss coefficient~which depends on the antenna characteristics,~and $\theta$ denotes the path-loss exponent.~Small scale fading is modeled by the Rayleigh fading.
\textcolor{blue}{We consider a cellular network in which a macro-cell is overlaid with four small-cells and 20 users and randomly distributed in the considered simulation region.~Unless specified,~we set  $\sigma^{2}=-120$~dBm,~$I^{m}_\textrm{th}=I_\textrm{th}=-90$~dBm,~$R_{\min}=5$~bps/Hz,~$A=2$,~and $M=8$.~Also,~the carrier
frequency and sub-channel bandwidth are 2 GHz and 180 kHz, respectively.}}

\begin{figure}[t] \label{figSC}
  \centering
  \includegraphics[height= 8.5 cm, width=10.500cm]{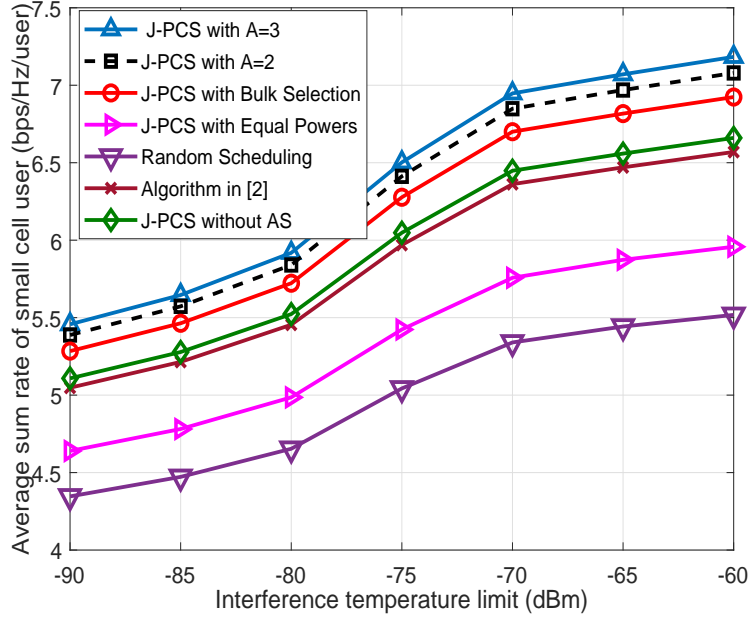}
  \vspace{-2mm}
  \small \caption{\textcolor{black}{\small Average sum rate of small cell users versus the interference
temperature limit~($I_{\mathrm{th}}$)~dBm.}}
\end{figure}

\textcolor{blue}{ Fig. 1 illustrates the spectral efficiency of small-cell users versus the interference threshold $I_{\mathrm{th}}$. For low interference threshold, constraint $C_2$ puts a stringent limitation on the maximum allocated power to small-cell users which results in low sum rate of small-cell users.~As the interference threshold increases, the allocated power for small-cell users increases which improves the average spectral efficiency per user. We observe the following important scenarios,~namely:
{\bf (i)} We investigate the performance of J-PCS algorithm (with per-subcarrier antenna selection) considering $A=2$ and $A=3$.~This setting is to facilitate the observation of spatial diversity gains when the number of antennas increases,~the number of independent paths between the users and BSs increases, thereby can be exploited to improve the system throughput;
 {\bf (ii)} We investigate the performance of J-PCS algorithm (with bulk antenna selection [8]) with $A=2$. The J-PCS algorithm with per-subcarrier antenna selection outperforms due to its ability in exploiting the frequency selective nature of the fading  channels as compared to the bulk selection\footnote{{Bulk antenna selection is considered by assuming that all sub-carriers can be assigned to one of the antennas, i.e., $\underset{m\in \mathcal{M}}{\overset{}{\mathop \sum }}\underset{a\in \mathcal{A}}{\overset{}{\mathop \sum }}\,x_{i}^{m a}=1$.}};
{\bf (iii)} equal power allocation (EPA) is generally considered as an efficient sub-optimal solution which can achieve a close-to-optimal performance in high SNR for single-cell networks when there is no interference. However, in our problem, EPA
does not achieve a good performance as it fails to harness the strong co-channel
interference. Thus, we adopt it as a benchmark to illustrate the performance gain of our proposed ALM power control,
 and {\bf (iv)} we compare  J-PCS without AS (J-PCS with A = 1) algorithm with the subchannel and power allocation algorithm presented in [2]. For the sake of  fair comparison, we assume a single antenna is available at the small-cell users. We note that  J-PCS algorithm can achieve a superior performance compared to the one in [2],~even if AS is not performed. The reason is due to the effectiveness of the proposed joint resource allocation and user association for better utilization of limited system resources.
}

{Fig. 2 is provided to show that impact of different values of $\mu_{1},\mu_{2}$ on spectral efficiency.~This figure also shows that for large penalty factors (greater than a threshold),~the objective function in (2) remains unchanged,~indicating that the penalty function approaches
to zero.~This confirms the claim that for $\mu_{i} \geq\mu^{*}$, the sub-channel-BS/antenna take binary values.}

In Fig. 3 we illustrate the overall convergence of our proposed iterative algorithm including the scheduling and power control.~These curves are obtained for different initial power allocations in which depend on starting point while all curves converge to almost the same value.~Moreover,~this figure shows that the sum rate is monotonically increasing at each iteration as it discussed in (3).~As can be observed, the algorithm quickly converges. \textcolor{black}{The figures also show that} less 20 iterations the performance of the proposed algorithm converges to a stationary point.
\begin{figure}
\centering
 \centering
  \includegraphics[height= 8.500cm,width=10.500cm]{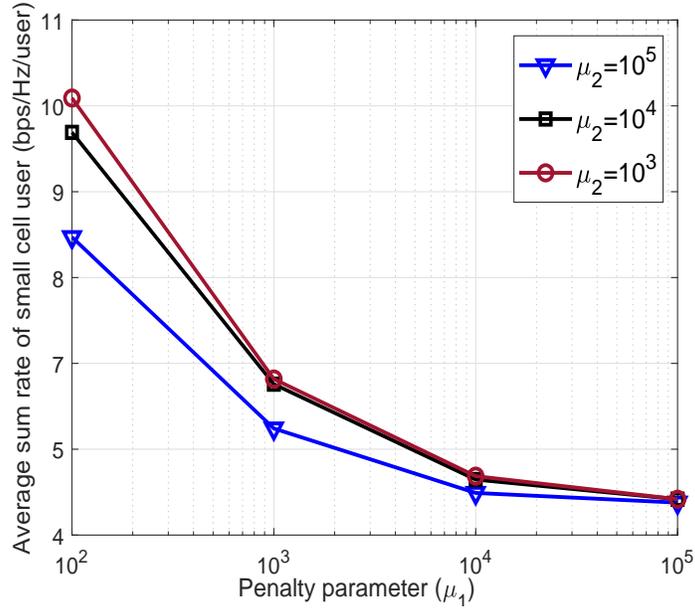}
  \small \caption{\small Effect of penalty parameters on the average small cell users.}
\label{fig:Penalty}
  \end{figure}
  \begin{figure}
  \centering
\includegraphics[width=10.500cm , height=8.5cm]{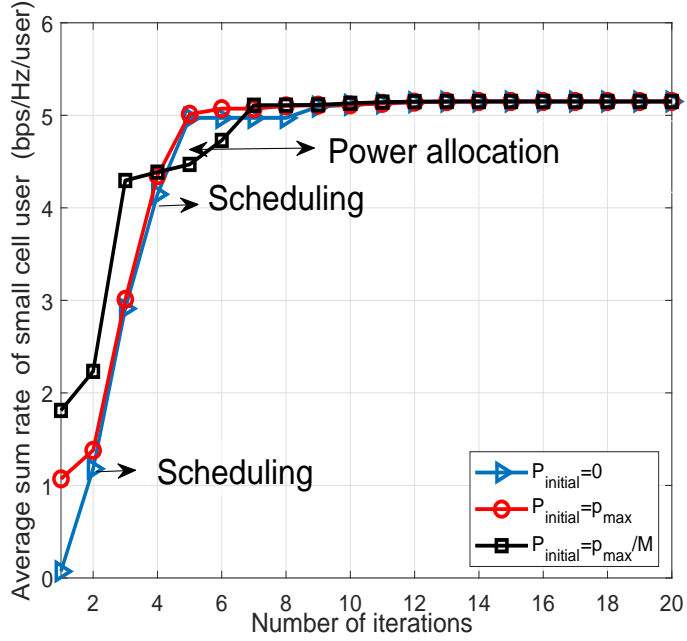}
  \small \caption{\small{Sum rate of small cell users versus the number of iterations for different initial power allocation.}}
  \label{fig:Comparision}
\end{figure}
\vspace{-4.0mm}
\section{Conclusion}\label{sec_conc}
\vspace{-2.0mm}
In this letter, a joint BS-sub-channel assignment,~AS and power control in the~UL~direction of a heterogeneous network are considered.  The optimization problem is decomposed into two sub-problems and each of the  subproblem was solved iteratively using tools from majorization-minimization and augmented Lagrangian method, respectively.~\textcolor{black}{Numerical results showed that proposed J-PCS algorithm outperforms the schemes addressed in the existing literature.}\textcolor{black}{~The extension to imperfect CSI scenarios will be considered in our future work.}
\vspace{-10.0mm}

\appendices
\section{}
\vspace{-3.0mm}
Let us consider the standard form an optimization problem:
\begin{align}
 \min_{\textbf{x}}~&f(\textbf{x})\nonumber\\
 \text{subject to}~~
 &h_{i}(\textbf{x})=0,~ \forall i \in\{1,...,m\},\nonumber\\
 &g_{j}(\textbf{x})<0,~ \forall j\in\{1,...,r\},\nonumber \tag{18}
 \end{align}
where $\textbf{x}$ is the optimization variable vector,~$f(\textbf{x})$ is the objective function,~$h_{i}$,~and $g_{j}$  denote the equality and inequality constraints,~respectively.
To solve the optimization problem (18),~we convert the inequality constraints $g_{j}(\textbf{x})<0$ to equality constraints as follows:
\begin{align}
g_{j}(\textbf{x})<0\rightarrow g_{j}(\textbf{x})+\mu^{2}_{j}=0,~\forall j=\{1,...,r\},\tag{19}
 \end{align}
It is worth mentioning that $\textbf{x}^{*}$ is a local minimum if and only if $(\textbf{x}^{*},\mu^{*}_{1},...,\mu^{*}_{r})$,~where $\mu^{*}_{j}=\sqrt{-g_{j}(\textbf{x}^{*})}$, $j=1,...,r$ is a local (global) minimum of problem (18).~Thus,~the augmented Lagrangian function can be expressed as follows:
\begin{align}
&\min_{\textbf{x},\mu}L_{\gamma}(\textbf{x},\mu,\eta,\lambda)=f(\textbf{x})+\eta h(\textbf{x})+\frac{\gamma}{2} ||h(\textbf{x})||^{2}\nonumber\\&+\sum_{j=1}^{r}\bigg\{  \lambda_{j}\bigg(g_{j}(\textbf{x})+\mu^{2}_{j}\bigg)+\frac{\gamma}{2}|g_{j}(\textbf{x})+\mu^{2}_{j}|^{2}\bigg\}.\tag{20}
 \end{align}
To resolve (20),~we first minimize $L_{\gamma}(\textbf{x},\mu,\eta,\lambda)$ with respect to $\mu$ as:
\begin{align}
&L_{\gamma}(\textbf{x},\eta,\lambda)=\min_{\mu}L_{\gamma}(\textbf{x},\mu,\eta,\lambda)=f(\textbf{x})+\eta h(\textbf{x})+\frac{\gamma}{2} ||h(\textbf{x})||^{2}\nonumber\\&+\sum_{j=1}^{r} \min_{\mu_{j}}\bigg\{  \lambda_{j}\bigg(g_{j}(\textbf{x})+\mu^{2}_{j}\bigg)+\frac{\gamma}{2}|g_{j}(\textbf{x})+\mu^{2}_{j}|^{2}\bigg\}.\tag{21}
 \end{align}
Then,~we minimize $L_{\gamma}(\textbf{x},\eta,\lambda)$ with respect to $\textbf{x}$ as follows:
\begin{align}
&L_{\gamma}(\textbf{x},\eta,\lambda)=f(\textbf{x})+\eta h(\textbf{x})+\frac{\gamma}{2} ||h(\textbf{x})||^{2}\nonumber\\&+\frac{1}{2\gamma}\Bigg(\sum_{j=1}^{r}\Big( \max\bigg\{0,  \lambda_{j}+\gamma g_{j}(\textbf{x})\bigg\}\Big)^{2}-\lambda_{j}^{2}\Bigg).\tag{22}
\end{align}
Furthermore,~the iterations for $\eta$ and $\lambda$ are implemented as:
\begin{align}
\eta^{*}_{i}\rightarrow \eta^{t}_{i}+\gamma^{t}h_{i}(x^{t}),
\lambda^{*}_{j}\rightarrow \max \bigg\{ 0,\lambda^{t}_{j}+\gamma^{t}g_{j}(x^{t})\bigg\}.\tag{23}
\end{align}

\begin{IEEEbiography}[{\includegraphics[width=1.1in,height=1.35in]{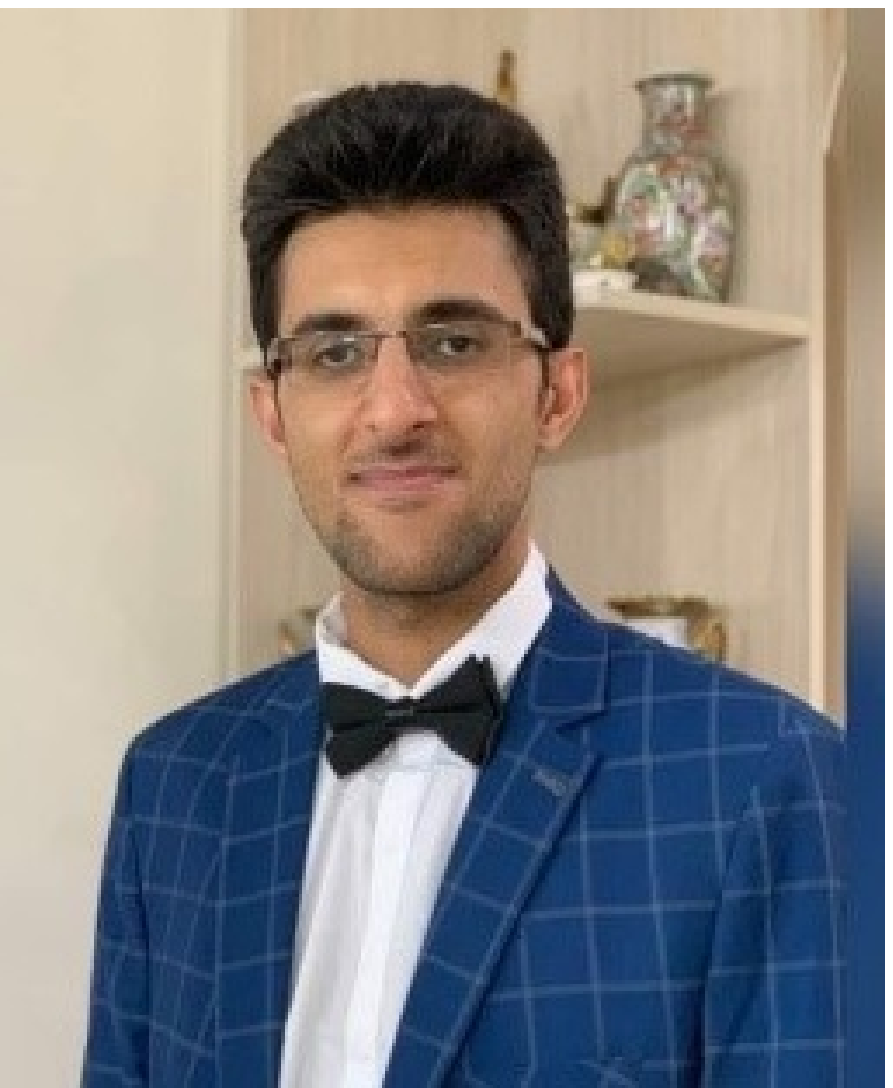}}]
{Ata Khalili} (S'18) received the B.Sc. degree and M.Sc. degree with first class honors in Electronic Engineering and Telecommunication Engineering from Shahed University in 2016 and 2018, respectively. He was a visiting researcher at the Department of Computer Engineering and Information Technology, Amirkabir University of Technology, Tehran, Iran.~Now he is working as a research assistant at the Department of Electrical and Computer Engineering Tarbiat Modares University,~Tehran,~Iran.~His research interests include convex and non-convex optimization, resource allocation in wireless communication, Green communication, and mobile edge computing.
\end{IEEEbiography}

\begin{IEEEbiography}[{\includegraphics[width=1.1in,height=1.35in]{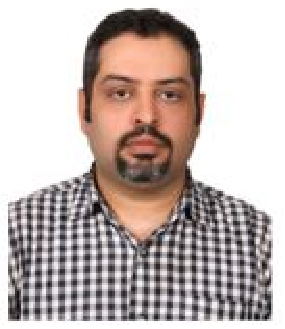}}]
{Soroush Akhlaghi} received the BASc degree in electrical engineering from K.N. Toosi University of Technology, Tehran, Iran, in 1999, the M.Sc. degree from Amirkabir University of Technology, Tehran, Iran, in 2001, and the Ph.D. degree from Iran University of Science and Technology (IUST), in 2007. In 2005, while working toward the Ph.D. degree, he joined the Department of Electrical and Computer Engineering, University of Waterloo, Canada, as a research assistant in the Coding and Signal Transmission Laboratory. He collaborated with this laboratory till the middle of 2010 in the area of new emerging wireless technologies. He is currently an assistant professor in the Engineering Department of Shahed University, Tehran, Iran. His research interests include multi-user communication networks, network coding, and optimization techniques for communication systems.
\end{IEEEbiography}

\begin{IEEEbiography}[{\includegraphics[width=1.1in,height=1.25in]{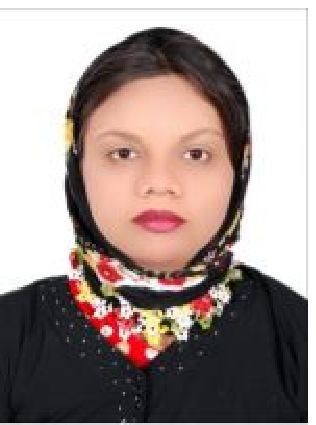}}]
{Hina Tabassum} (SM'17) is an Assistant Professor at the Lassonde School of Engineering, York University, Canada. She received the B.E. degree in electronic engineering from the NED University of
Engineering and Technology (NEDUET), Karachi,
Pakistan, in 2004. She received during her undergraduate studies two gold medals from NEDUET and SIEMENS for securing the first position among all engineering universities of Karachi. She then
worked as lecturer in NEDUET for two years. In September 2005, she joined the Pakistan Space and Upper Atmosphere Research Commission (SUPARCO), Karachi, Pakistan and received there the best performance award in 2009. She completed her Masters and Ph.D. degree in Communications Engineering from NEDUET in
2009 and King Abdullah University of Science and Technology (KAUST), Saudi Arabia, in May 2013, respectively. From 2013 till June 2018, she was a research fellow at the Department of Electrical and Computer Engineering, University of Manitoba, Canada. Her research interests include 5G wireless
networks with focus on their stochastic geometry-based performance modeling and optimization, internet-of-things (IoT), and machine learning applications. She is a registered Professional Engineer in the province of Manitoba, Canada.
\end{IEEEbiography}

\begin{IEEEbiography}[{\includegraphics[width=0.95in,height=1.2in]{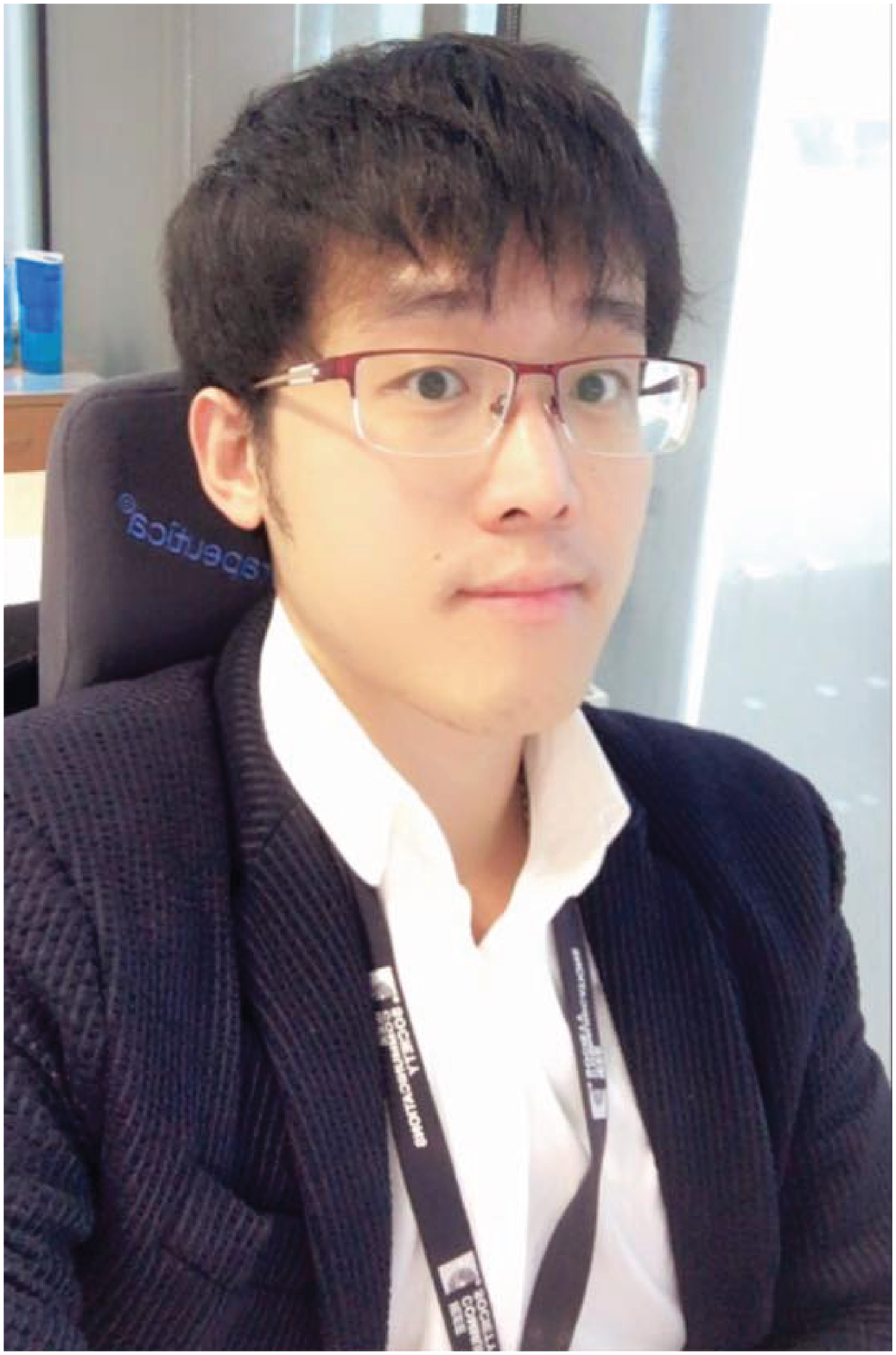}}]{Derrick Wing Kwan Ng} (S'06-M'12-SM'17)   received the bachelor degree with first-class honors and the Master of Philosophy (M.Phil.) degree in electronic engineering from the Hong Kong University of Science and Technology (HKUST) in 2006 and 2008, respectively. He received his Ph.D. degree from the University of British Columbia (UBC) in 2012. He was a senior postdoctoral fellow at the Institute for Digital Communications, Friedrich-Alexander-University Erlangen-N\"urnberg (FAU), Germany. He is now working as a Senior Lecturer and an ARC DECRA Research Fellow at the University of New South Wales, Sydney, Australia.  His research interests include convex and non-convex optimization, physical layer security, wireless information and power transfer, and green (energy-efficient) wireless communications.
Dr. Ng received the Best Paper Awards at the IEEE TCGCC Best Journal Paper Award 2018, INISCOM 2018, IEEE International Conference on Communications (ICC) 2018,  IEEE International Conference on Computing, Networking and Communications (ICNC) 2016,  IEEE Wireless Communications and Networking Conference (WCNC) 2012, the IEEE Global Telecommunication Conference (Globecom) 2011, and the IEEE Third International Conference on Communications and Networking in China 2008.  He has been serving as an editorial assistant to the Editor-in-Chief of the IEEE Transactions on Communications since Jan. 2012. In addition, he is listed as a Highly Cited Researcher by Clarivate Analytics in 2018 and 2019.
\end{IEEEbiography}

\vspace{-9mm}
\begin{thebibliography}{1}
\vspace{-3mm}
\bibitem{Dr.Tabassum}U. Siddique, H. Tabassum, E. Hossain, and D. I. Kim, ``Channel-access-aware user association with interference coordination in two-tier downlink cellular networks," \textit{IEEE Trans.~Veh Technol,} vol. 65, no. 7, pp. 5579-5594, Jul.~2016.
 \bibitem{3} B. Khamidehi, A. Rahmati, and M. Sabbaghian, ``Joint sub-channel assignment and power allocation in heterogeneous networks: An efficient optimization method," \textit{IEEE Commun. Lett.,} vol. 20, no. 12, pp. 2490-2493, Dec. 2016.
    
\bibitem{multi} F. Wang, W. Chen, H. Tang, and Q. Wu, ``Joint optimization of user association, subchannel allocation, and power allocation in multi-cell multi-association OFDMA heterogeneous networks," \textit{IEEE Trans. Commun.,} vol. 65, no. 6, pp. 2672-2684, Jun.~2017.

\bibitem{massive}H. Ma, H. Zhang, X. Wang, and J. Cheng,``Backhaul-aware user association and resource allocation for massive MIMO-enabled HetNets," \textit{IEEE Commun.~Lett,} vol. 21, no. 12, pp. 2710-2713, Dec. 2017.
\textcolor{black}{\bibitem{Poor} H. Zhang, S. Huang, C. Jiang, K. Long, V. C. M. Leung, and H. V. Poor, ``Energy efficient user association and power allocation in millimeter-wave-based ultra dense networks with energy harvesting base stations," \textit{IEEE J. Sel. Areas Commun.,} vol. 35, no. 9, pp. 1936-1947, Sep. 2017}
\textcolor{blue}{\bibitem{Contract}J. Du, E. Gelenbe, C. Jiang, H. Zhang, and Y. Ren, ``Contract design for traffic offloading and resource allocation in heterogeneous ultra-dense networks," \textit{IEEE J. Sel. Areas Commun.,} vol. 35, no. 11, pp. 2457-2467,~Nov. 2017.}


 \bibitem{globecom2018}U. Bin Farooq, U. Sajid Hashmi, J. Qadir, A. Imran, and A. N. Mian ``User transmit power minimization through uplink resource allocation and user association in HetNets" \textit{IEEE Proc.~Globecom}, pp. 1-6.~Dec.~2018 
\textcolor{blue}{\bibitem{LTE}N. P. Le, F. Safaei, and L. C. Tran, ``Antenna selection strategies for MIMO-OFDM wireless systems: An energy efficiency perspective,”
\textit{IEEE Trans. Veh. Technol.,} vol. 65, no. 4, pp. 2048-2062, Apr. 2016.}
\bibitem{Beyound} J. Zhang, E. Björnson, M. Matthaiou, D. W. K. Ng, H. Yang, and D.J. Love, ``Multiple Antenna Technologies for Beyond 5G". \textit{arXiv:1910.00092.}
    \bibitem{Dr. Kwan}D. W. K. Ng, Y. Wu, and R. Schober, ``Power efficient resource allocation for full-duplex radio distributed antenna networks," \textit{IEEE Trans. Wireless Commun.}, vol. 15, no. 4, pp. 2896-2911, Apr. 2016.
\bibitem{10}A. Khalili, S. Zarandi, and M. Rasti, ``Joint resource allocation and offloading decision in mobile edge computing," \textit{IEEE Commun. Lett.,}~vol. 23, no. 4, pp. 684-687, Apr.~2019.
    \bibitem{11}Y. Sun, P. Babu, and D. P. Palomar, ``Majorization-Minimization algorithms in signal processing, communications, and machine learning,"
\textit{IEEE Trans. Signal Process.,} vol. 65, no. 3, pp. 794-816, Feb. 2017.
    \bibitem{12}D. P. Berteskas, \textit{Nonlinear Programming: 2nd Edition.} MIT Press, 1999.
    \bibitem{Boyd}S. Boyd and L. Vandenberghe, \textit{Convex Optimization.} Cambridge
University Press, 2004.
\textcolor{black}{\bibitem{Path} Ericsson, ST-Ericsson, ``Uplink system-level simulation results in cochannel deployment with full buffer traffic (R1-131537)," Apr. 2013, 3GPP TSG RAN WG1 Meeting-72bis.}
\end{thebibliography}
\end{document}